\documentclass[prl,superscriptaddress,showpacs,longbibliography,reprint]{revtex4-2} 

\usepackage{hyperref}
\usepackage{color}
\usepackage{stmaryrd}

\usepackage[usenames,dvipsnames]{xcolor}
\usepackage{amsmath}
\usepackage{amssymb}
\usepackage{graphicx}
\graphicspath{{graphics/}}
\usepackage{epsfig}
\usepackage{dcolumn}
\usepackage{bm}
\usepackage{mathrsfs}
\usepackage{multirow}
\usepackage[all]{xy}
\usepackage{pbox}
\usepackage{lipsum}
\usepackage{verbatim}
\usepackage{braket}
\usepackage{dsfont}
\usepackage{array}
\usepackage{makecell}
\usepackage{tabularx}
\usepackage{isotope}

\usepackage{color,soul}

\begin{document}
\title{Numerical simulations of long-range open quantum many-body dynamics with tree tensor networks}

\author{Dominik Sulz}
\affiliation{Mathematisches Institut, Universit\"at T\"ubingen, Auf der Morgenstelle 10, D–72076 T\"ubingen, Germany}

\author{Christian Lubich}
\affiliation{Mathematisches Institut, Universit\"at T\"ubingen, Auf der Morgenstelle 10, D–72076 T\"ubingen, Germany}

\author{Gianluca Ceruti}
\affiliation{Institute of Mathematics, EPF Lausanne, 1015 Lausanne, Switzerland}

\author{Igor Lesanovsky}
\affiliation{Institut für Theoretische Physik, Universit\"at T\"ubingen, Auf der Morgenstelle 14, 72076 T\"ubingen, Germany}
\affiliation{School of Physics and Astronomy and Centre for the Mathematics and Theoretical Physics of Quantum Non-Equilibrium Systems, The University of Nottingham, Nottingham, NG7 2RD, United Kingdom}

\author{Federico Carollo}
\affiliation{Institut für Theoretische Physik, Universit\"at T\"ubingen, Auf der Morgenstelle 14, 72076 T\"ubingen, Germany}

\begin{abstract}
Open quantum systems provide a conceptually simple setting for the exploration of collective behavior stemming from the competition between quantum effects, many-body interactions, and dissipative processes. They may display dynamics distinct from that of closed quantum systems or undergo nonequilibrium phase transitions which are not possible in classical settings. However, studying open quantum many-body dynamics is challenging, in particular in the presence of critical long-range correlations or long-range interactions. Here, we make progress in this direction and introduce a numerical method for open quantum systems, based on tree tensor networks. Such a structure is expected to improve the encoding of many-body correlations and we adopt an integration scheme suited for long-range interactions and applications to dissipative dynamics. We test the method using a dissipative Ising model with power-law decaying interactions and observe signatures of a first-order phase transition for power-law exponents smaller than one.
\end{abstract}

\maketitle

The interaction of a quantum system with its surroundings induces dissipative effects which require the description of its state in terms of density matrices. In the simplest case, these matrices evolve through Markovian quantum master equations  \cite{lindblad1976,gorini1976,breuer2002}. However, solving these equations for many-body systems is a daunting task, especially beyond noninteracting theories \cite{prosen2008,prosen2010,prosen2010b,heinosaari2010,guo2017}. This is due to the exponential growth (with the system size) of the resources needed to encode quantum states, which seriously limits the investigation of nonequilibrium behavior in open quantum systems  \cite{dagvadorj2015,roscher2018,dogra2019,buca2019,chiacchio2019,carollo2019,jo2021,carollo2022,helmrich2020,diehl2008,verstraete2009,weimer2010}.
\begin{figure}[t]
\centering
\includegraphics[width=\columnwidth]{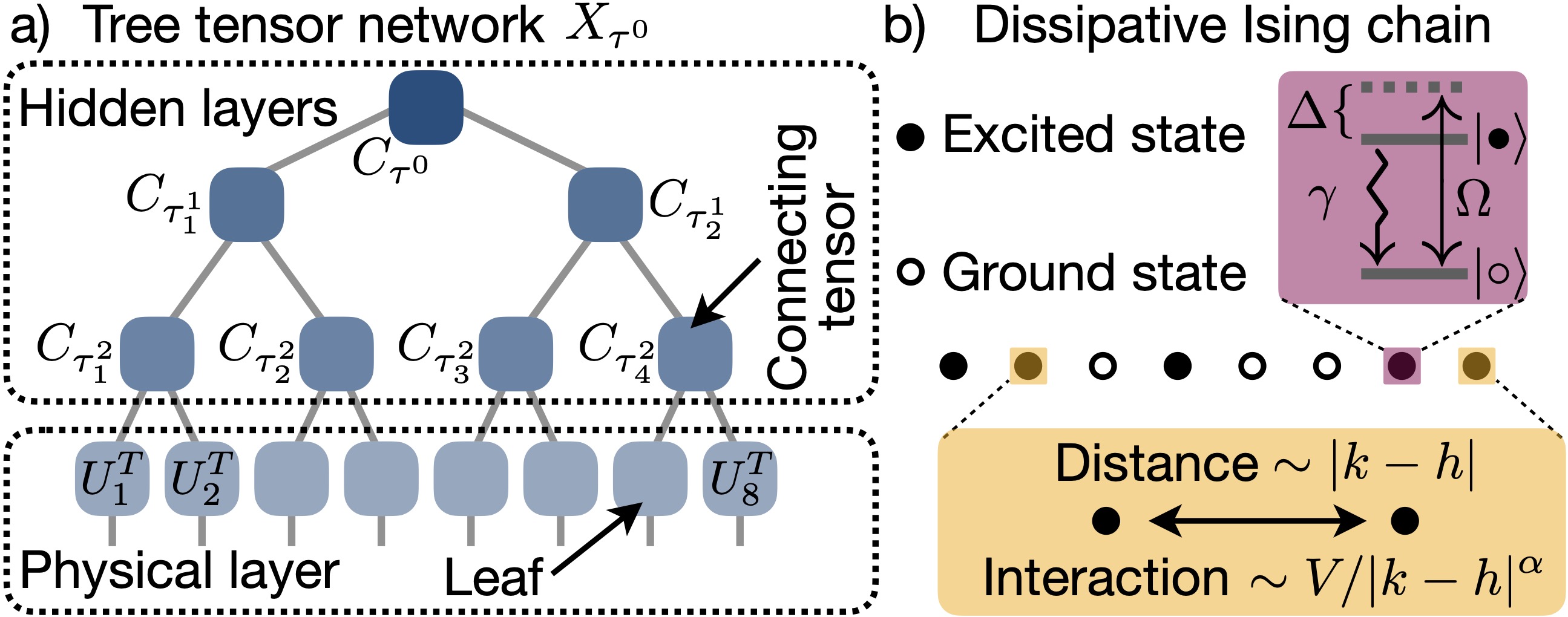}
\caption{{\bf Tree tensor networks and dissipative Ising model.} (a) Binary tree tensor network $X_{\tau^0}$ living on a tree $\tau^0$, associated with a system made by $D=8$ particles. The network consists of a physical layer, containing the leaves of the tree (each one related to a particle), and three hidden layers. The latter are made by connecting tensors $C_{\tau_{i}^j}$, pairwise joining elements from the previous layer. (b) Dissipative Ising model with two-level subsystems and single-particle states $\ket{\bullet}$, $\ket{\circ}$. The system Hamiltonian drives coherent oscillations between states  $\ket{\bullet}\leftrightarrow \ket{\circ}$ with Rabi frequency $\Omega$ and detuning $\Delta$. The irreversible process consists of local decay $\ket{\bullet}\rightarrow \ket{\circ}$ with rate $\gamma$. Two subsystems in state $\ket{\bullet}$ interact with a strength depending on the parameter $V$ and on their distance through the interaction-range exponent $\alpha$.}
\label{Fig1}
\end{figure} 

To overcome this limitation, several numerical approaches have been developed \cite{cui2015,jin2016,werner2016,jaschke2018,silvi2019,weimer2021}, including techniques based on neural networks \cite{yoshioka2019,hartmann2019,nagy2019,vicentini2019,reh2021}. At least for one-dimensional quantum systems, the state-of-the-art methodology is based on matrix product states (MPSs) \cite{vidal2003,vidal2004,schollwock2011,orus2014,LuOV2015,haegeman2016,paeckel2019,orus2019,cirac2021}, despite open questions on their performance for open quantum dynamics \cite{werner2016} and on error bounds for the estimation of expectation values. These aspects are particularly relevant close to nonequilibrium  phase transitions, where MPS methods can become unstable \cite{carollo2019,carollo2022}, since they struggle to capture long-range correlations in critical systems or in systems with long-range interactions.

Recently, tree tensor networks (TTNs), which are tensor networks featuring both a physical and several hidden layers [see sketch in Fig.~\ref{Fig1}(a)], have been successfully employed to encode critical long-range correlations \cite{silvi2010} in Hamiltonian systems \cite{shi2006,nakatani2013,schroder2019,arceci2022} (see also Refs.~\cite{tagliacozzo2009,murg2010,kloss2020,felser2021} for other applications). This enhanced capability is rooted in their structure  [cf.~Fig.~\ref{Fig1}(a)], which is such that the number of tensors between two subsystems scales only logarithmically with their distance \cite{shi2006} and not linearly as for MPSs. Despite this feature, TTNs have not yet been used for simulating critical or long-range open quantum dynamics \cite{weimer2021} (see, however, related ideas in Refs.~\cite{finazzi2015,rota2017,rota2019}). 

In this paper, we present an algorithm
for simulating quantum master equations which exploits a TTN representation of quantum many-body states. Our approach, based on the integration scheme put forward in Ref.~\cite{CLS2022}, evolves a TTN  by a hierarchical ``basis update \& Galerkin'' (BUG) method. It first 
updates the orthonormal basis matrices which are found at the {\it leaves} of the tree  and then evolves the {\it connecting tensors} within the hidden layers [see Fig.~\ref{Fig1}(a)], by a variational, or Galerkin, method.  

To benchmark our algorithm, we consider a paradigmatic open quantum system, the dissipative Ising model sketched in Fig.~\ref{Fig1}(b), in the presence of power-law decaying interactions. We show the validity of the method by checking it against (numerically) exact results for both short-range and long-range interactions and we investigate signatures of a first-order phase transition in the long-range scenario. We further consider a global ``susceptibility" observable and explore how TTNs perform in describing many-body correlations. Our results indicate that TTNs are
promising for simulating open quantum many-body systems in the presence of long-range interactions. 
\\

\noindent {\bf Open quantum dynamics.---} We consider one-dimensional quantum systems consisting of $D$ distinguishable $d$-level particles undergoing Markovian open quantum dynamics. The density matrix $\rho(t)$  describing the state of the system evolves through the quantum master equation \cite{lindblad1976,gorini1976,breuer2002}
\begin{equation}
\dot{\rho}(t)=\mathcal{L}[\rho(t)]:=-i[H,\rho(t)]+\mathcal{D}[\rho(t)]\, .
\label{QME}
\end{equation}
The map $\mathcal{L}$ is the Lindblad dynamical generator and $H=H^\dagger$ is the many-body Hamiltonian operator. 
The dissipator $\mathcal{D}$ assumes the form 
\begin{equation}
\mathcal{D}[\rho]=\sum_{\mu}\left(J_\mu \rho J_\mu^{\dagger}-\frac{1}{2}\left\{\rho, J_\mu^\dagger J_\mu \right\}\right)\, ,
\label{D}
\end{equation}
with the jump operators $J_\mu$ encoding how the environment affects the system dynamics. 

The Lindblad generator in Eq.~\eqref{QME} is a linear map from the space of matrices onto itself. To numerically simulate open quantum dynamics, it is convenient to represent  $\mathcal{L}$ as a matrix acting on a vectorized representation of matrices
(see e.g.~\cite{verstraete2004,zwolak2004,kshetrimayum2017,carollo2022}).
Any matrix $\rho(t)$ thus becomes a vector $\ket{\rho(t)}$, and Eq.~\eqref{QME} reads 
\begin{equation}
    \ket{\dot{\rho}(t)}=\mathbb{L}\ket{\rho(t)}\, ,
    \label{vec_QME}
\end{equation}
with $\mathbb{L}$ being the matrix representation of the generator $\mathcal{L}$ (see Supplemental Material \cite{SM}\vphantom{\cite{KT2014}} for an example). 
In what follows, we show how the solution of Eq.~\eqref{vec_QME} can be approximated by means of TTNs. \\

\noindent {\bf Integration with tree tensor networks.---} 
TTNs feature a physical layer and several hidden layers [cf.~Fig.~\ref{Fig1}(a)], which we exploit to store physical data corresponding to the quantum state $\ket{\rho(t)}$. The leaves of the tree, i.e., the tensors (in fact,  matrices) in the physical layer, correspond to the sites of our one-dimensional quantum system while the connecting tensors in the hidden layers encode correlations between them. To approximate the open quantum dynamics, we adopt an algorithm \cite{CLS2022} that decomposes the 
Dirac--Frenkel time-dependent variational principle \cite{kramer1981,lubich2008,haegeman2011} for TTNs
into computable discrete time steps, variationally evolving each tensor of the TTN in a hierarchical order from the leaves to the root (bottom-up).
A single update of the algorithm consists of two steps,
\begin{enumerate}
    \item Construct a state-dependent reduction $L_{\tau_i^j}$ of the Lindblad generator, for each tensor in layer $j$,
    \item Update the tensor variationally, by solving the system of differential equations implemented by $L_{\tau_i^j}$, 
\end{enumerate}
which are repeated recursively going from the bottom to the top layer. We now provide a concise description of the algorithm and refer to Ref.~\cite{CLS2022} for details.

\begin{figure*}[t]
\centering
\includegraphics[width=\textwidth]{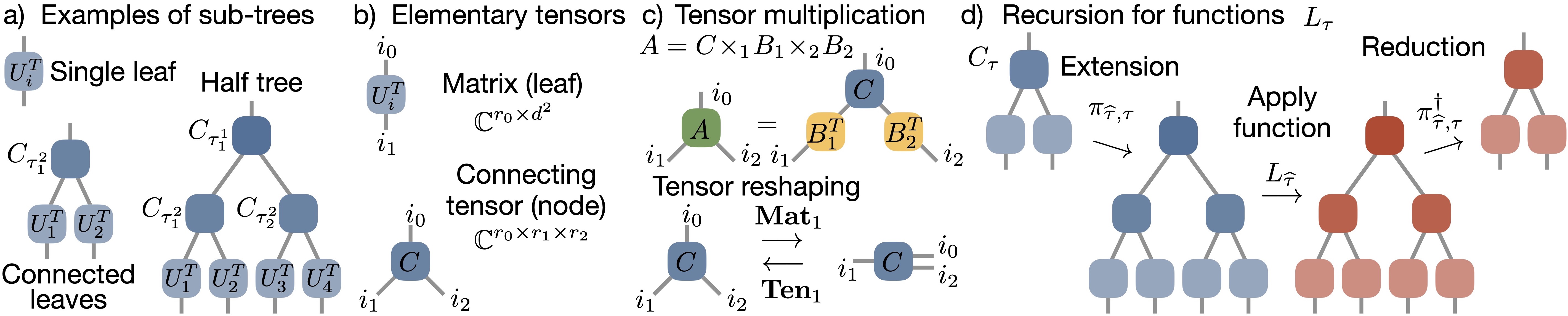}
\caption{{\bf Tree tensor networks: sub-trees and operations.} (a) Examples of tensor networks $X_{{\tau}}$, defined on sub-trees of the TTN shown in Fig.~\ref{Fig1}(a). Apart from the leaves, each $X_{{\tau}}$ has a connecting tensor $C_\tau$ at the top. (b) Leaves are complex orthogonal matrices with dimension $r_0 \times d^2$, where $d$ is the dimension of the single-particle  Hilbert space. Connecting tensors are order-three tensors with bond dimension $r_0\times r_1 \times r_2$. The indexes of the tensors are numbered from $0$ to $M-1$, where $M$ is the order of the tensor. (c) Sketch of the tensor multiplication [cf.~Eq.~\eqref{T-Mproduct}]  and of the reshaping functions ${\rm Mat}_i$, ${\rm Ten}_i$, shown for $i=1$. (d) Recursion algorithm for finding ${L}_{{\tau}}$. Given  ${L}_{\widehat{\tau}}$, with $\widehat{\tau}$ being the smallest tree larger than ${\tau}$ and containing it, to find ${L}_{{\tau}}$ we extend the tensor network $X_{{\tau}}$ 
to a tensor network on the larger tree 
$\widehat{\tau}$, via the operation $\pi_{\widehat{\tau},{\tau}}$. We apply the function ${L}_{\widehat{\tau}}$ on the extended tensor network and then reduce the resulting tensor back to ${\tau}$ through  $\pi_{\widehat{\tau},{{\tau}}}^\dagger$.}
\label{Fig2}
\end{figure*}

The physical layer is formed by $D=2^\ell$ leaves, each one associated with a site of the system. 
Each leaf is the smallest possible sub-tree, which we call $\tau^\ell_{i}$, see Fig.~\ref{Fig2}(a). The superscript $\ell$ labels the (physical) layer to which the leaf belongs, while $i$, for $i=1,2\dots D$, denotes the leaf itself.
The tensor associated with each leaf is a complex matrix $U_i$, with orthonormal columns and dimensions ${d^2\times r_{\tau^\ell_i}}$, carrying a physical basis. We will use the letter $r$ to denote bond dimensions within the TTN. 

Proceeding towards the root (top) of the tree depicted in Fig.~\ref{Fig1}(a), for each hidden layer $j$ we can recursively define larger sub-trees $\tau^{j}_{i}$ obtained by joining sub-trees from the previous layer, namely $\tau^{j}_{i}=(\tau^{j+1}_{2i-1} ,\tau^{j+1}_{2i})$ [cf.~Fig.~\ref{Fig2}(a)]. Each sub-tree is related to a tensor network $X_{\tau^{j}_{i}}$, at the root of which one finds the connecting tensor $C_{\tau^j_{i}}$, with dimension  $r_{\tau^j_i}\times r_{\tau^{j+1}_{2i-1}}\times r_{\tau^{j+1}_{2i}}$.
As shown in Fig.~\ref{Fig2}(b), the first index of these tensors (bond dimension $r_{\tau^j_i}$) points upward and is counted as the zeroth dimension, followed by the second and third indeces (bond dimensions $r_{\tau^{j+1}_{2i-1}}$ and $r_{\tau^{j+1}_{2i}}$) which point downward to the left and right sub-tree, respectively.

We further define the tensor-matrix multiplication $A=C \times_{m}B$, between an order-$n$ tensor $C$ and a matrix $B$ with respect to $m$th tensor index as [see, e.g., Fig.~\ref{Fig2}(c)]
\begin{equation}
A_{k_0,k_1,\dots k_m\dots  k_{n-1}}=\sum_{\ell_m}C_{k_0,k_1\dots \ell_m \dots k_{n-1}}(B^T)_{\ell_m,k_m}\, ,
    \label{T-Mproduct}
\end{equation}
as well as the {\it matricization} of a tensor ${\rm Mat}_i(C)=C_i\in\mathbb{C}^{r_i \times r'_i}$, where $r'_i=\prod_{j\neq i}r_j$, with inverse operation, ${\rm Ten}_i(C_i)=C$, called {\it tensorization} [cf.~Fig.~\ref{Fig2}(c)]. We work with orthonormal TTNs, for which ${\rm Mat}_0(C)^T$ has orthonormal columns for each connecting tensor $C$, with the exception of the connecting tensor $C_{\tau^0}$ at the root (top). The relation $r_i\le \prod_{j\neq i}r_j$ must be satisfied for $i=0,1,2$, to ensure that
each matricization of each connecting tensor $C$ can be (and usually is) of full rank.

To obtain the evolved TTN over a discrete (infinitesimal) time-step $\delta t$, we need to find the updated leaves $U_i'$ and the updated connecting tensors $C_{\tau_i^j}'$. In the algorithm, we first update the basis matrices $U_i$ at the leaves. To this end, we take the connecting tensor above the leaf $\tau^\ell_i$ that needs to be updated, $C=C_{\tau^{\ell-1}_{\lceil i/2 \rceil}}$, matricize it as $C_1={\rm Mat}_1(C)$ if $i$ is odd or as $C_2={\rm Mat}_2(C)$ if $i$ is even. We then perform a QR decomposition $C_i^T=Q_iR_i$, with $R_i$ having dimension $r_i\times r_i$, and define the matrix $Y_i=R_iU_i^T$. This provides the initial condition, $Y_i(0)=Y_i$, for the matrix differential equation 
\begin{equation}
\dot{Y}_i={L}_{\tau_i^\ell}[Y_i]\, .
\label{ODE-leaf}
\end{equation}
The linear operator ${L}_{\tau_i^\ell}$ can be interpreted as a state-dependent variational  reduction to the $i$th physical site of the Lindblad operator and, as we discuss below, is defined recursively. Solving the differential equation up to time $\delta t$, we find $Y_i(\delta t)$ and set the updated leaf matrix $U_i'$ as the orthogonal part of the QR factorization of $Y_i(\delta t)$.

We then hierarchically update the connecting tensors from  bottom to top layer. At each step of the recursion, we set $\widehat{C}_{\tau^j_i}=C_{\tau^j_i}\times_{1} M_{\tau_{2i-1}^{j+1}}\times_{2} M_{\tau_{2i}^{j+1}}$ [cf.~Fig.~\ref{Fig2}(c)], with $M_{\tau}=U'^{ \dagger}_{\tau}U_{\tau}$, where $U_{\tau}={\rm Mat}_0(X_{\tau})^T$ is the matricization of $X_\tau$. When $\tau$ is a leaf, $\tau=\tau_i^\ell$, then $U_{\tau_i^\ell}= U_i$.  The matrices $U'_{\tau}$ are defined analogously but for the already updated $X'_\tau$ \cite{SM}. The tensor $\widehat{C}_{\tau^j_i}$ provides the initial data for the differential equation 
\begin{equation}
\dot{\widehat{C}}_{\tau^j_i}={L}_{\tau_i^j}[\widehat{C}_{\tau^j_i}\times_{1} U'_{\tau_{2i-1}^{j+1}}\times_{2} U'_{\tau_{2i}^{j+1}} ]\times_{1} U_{\tau_{2i-1}^{j+1}}'^{ \dagger}\times_{2} U_{\tau_{2i}^{j+1}}'^{ \dagger}.
\label{ODE-C}
\end{equation}
The updated tensor is $C'_{\tau_{i}^{j}}={\rm Ten}_0 (Q^T)$, where $Q$ is the orthogonal part of the QR decomposition of $ {\rm Mat}_0 ( \widehat{C}_{\tau_{i}^{j}}(\delta t))^T$.
Note that the matrices $U_{\tau},U_\tau'$ are  never explicitly constructed since products involving them are computed by contracting corresponding TTNs. 

Finally, we discuss how the reduced Lindblad operators ${L}_{\tau_i^j}$ can be obtained  (see Refs.~\cite{CLW2021,SM} for details). For any sub-tree ${\tau}$, ${L}_{{\tau}}$ can be found from the knowledge of ${L}_{\widehat{\tau}}$ associated with the smallest sub-tree $\widehat{\tau}$ containing ${\tau}$. By defining a state-dependent extension operator $\pi_{\widehat{\tau},{\tau}}$, which maps the tree ${\tau}$ into the larger tree $\widehat{\tau}$, the operator ${L}_{{\tau}}$ is given by ${L}_{{\tau}}=\pi_{\widehat{\tau},{\tau}}^\dagger  {L}_{\widehat{\tau}} \pi_{\widehat{\tau},{\tau}}$. The starting point of the recursion is given by ${L}_{\tau^0}$, related to the whole tree $\tau^0$, which is nothing but a (possibly truncated) TTN-operator  representation of the matrix $\mathbb{L}$. 

The integrator presented above \cite{CLS2022}, which extends the BUG matrix integrators of Refs.~\cite{CL2021,CKL2022}, does not have any backward-in-time propagation in contrast to those of Refs.~\cite{LuOV2015,haegeman2016,CLW2021}. This makes it better suited for the simulation of dissipative dynamics. 
\\

\begin{figure}[t]
\centering
\includegraphics[width=\columnwidth]{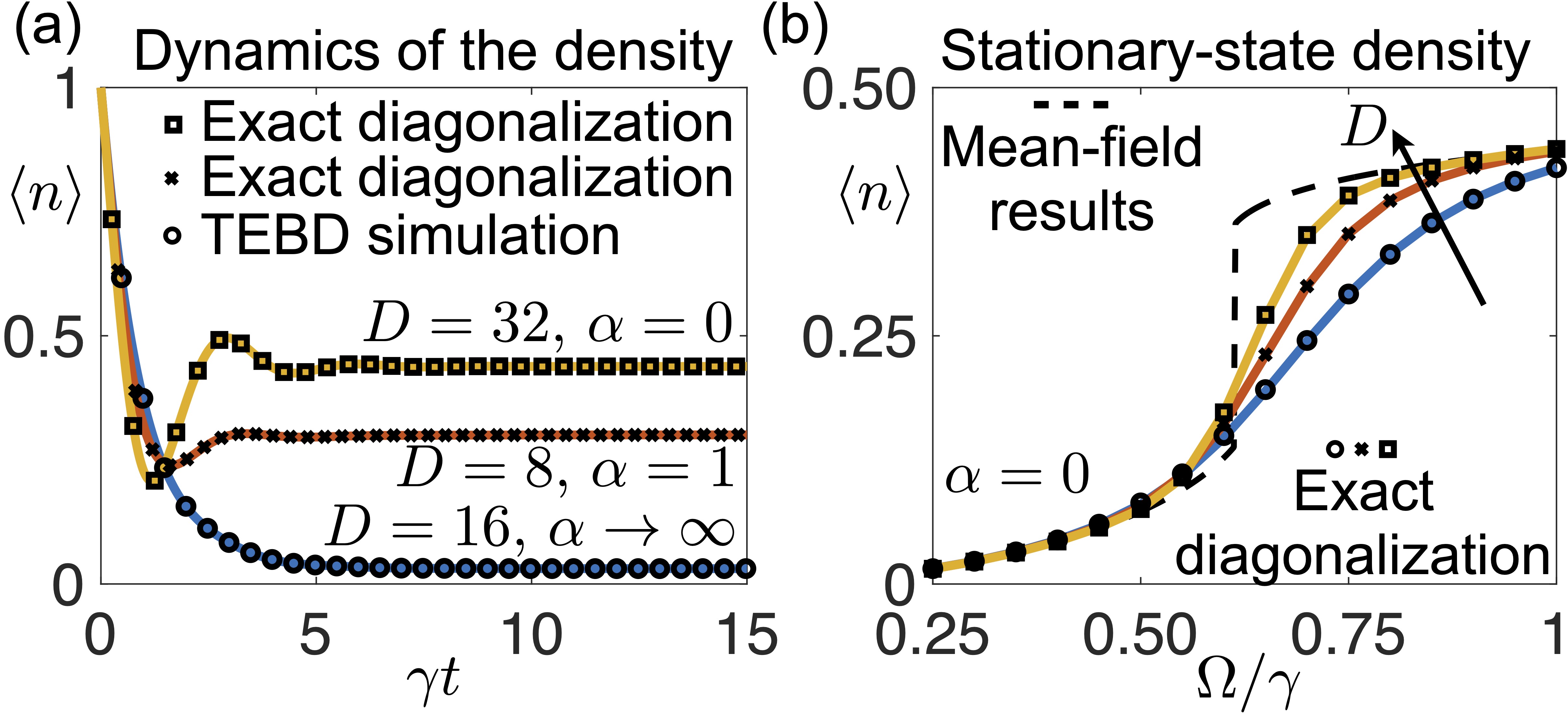}
\caption{{\bf Benchmark of the method.} (a) Time evolution of the density $\langle n\rangle$ computed with our algorithm (solid lines) and other approaches (symbols). For the nearest-neighbor interacting case ($D=16,\, \alpha\to\infty$), we benchmark our results against converged MPS simulations with a time-evolving-block-decimation (TEBD) algorithm.  For $D=8,\, \alpha=1$ and $D=32,\, \alpha=0$, we compare with results obtained with exact diagonalization of the Lindblad generator. For our simulations we used a maximal bond-dimension $r_{\rm max}=30$. We consider $\Omega/\gamma=0.3,0.7,1$ respectively for the different curves. (b) Stationary values of the density as a function of $\Omega/\gamma$, obtained by evolving up to $\gamma t=15$ the TTNs and exact-diagonalization simulations. Here, we show the infinite-range interacting case ($\alpha=0$), for system sizes $D=8,16,32$, for which we used maximal bond dimension $r_{\rm max}=10,20,30$, respectively. The dashed-line shows the mean-field prediction, which is exact for $D\to\infty$ \cite{benatti2018,carollo2021}. The parameters not explicitly specified in the panels are $\Delta/\gamma=-2$, $V/\gamma=5$.}
\label{Fig3}
\end{figure} 

\noindent {\bf Long-range dissipative Ising model.---}  To benchmark our algorithm, we consider a long-range interacting version of the  dissipative Ising model \cite{lee2011,weimer2015,cui2015,overbeck2017,raghunandan2018,jin2018,paz2021,Paz2021b}. It consists of a one-dimensional model with two-level particles, characterized by excited state $\ket{\bullet}$ and ground state $\ket{\circ}$. The model Hamiltonian [cf.~Fig.~\ref{Fig1}(b)] is given by 
\begin{equation}
H=\Omega \sum_{k=1}^D\sigma_x^{(k)}+\Delta \sum_{k=1}^D n^{(k)}+\frac{V}{2c_\alpha}\sum_{k\neq h =1}^D \frac{n^{(k)}n^{(h)}}{|k-h|^\alpha}\, ,
\label{H_2}
\end{equation}
where $n=\ket{\bullet}\!\bra{\bullet}$ and $\sigma_x=\ket{\bullet}\!\bra{\circ}+\ket{\circ}\!\bra{\bullet}$. The first two terms in the above equation describe a driving term, e.g., from a laser, with Rabi frequency $\Omega$ and detuning $\Delta$. The last term describes two-body interactions solely occurring between particles in the excited state $\ket{\bullet}$. The parameter $V$ is an overall coupling strength while the algebraic exponent $\alpha$ controls the range of the interactions. For $\alpha=0$ the interaction is of all-to-all type while for $\alpha\to\infty$ it only involves nearest neighbors. The coefficient $c_\alpha=\sum_{k=1}^D 1/k^\alpha$ keeps the interaction extensive for any value of $\alpha$ \cite{kac1963,defenu2021}. Dissipation  [cf.~Eq.~\eqref{D}] is encoded in the jump operators $J_k=\sqrt{\gamma}\sigma_-^{(k)}$, where $\sigma_-=\ket{\circ}\!\bra{\bullet}$ describes irreversible decay  from state $\ket{\bullet}$ to state $\ket{\circ}$. In the following, we shall consider as initial state the state with all particles in $\ket{\bullet}$. 

To show that our algorithm faithfully approximates the  open quantum dynamics, we test our results for different values of $\alpha$. When $\alpha=0$, we check our numerics against an efficient diagonalization method for the generator, possible for permutation-invariant systems  \cite{chase2008,baragiola2010,kirton2017,shammah2018}. For $\alpha\to\infty$, we compare numerical results with those obtained using MPSs and a time-evolving-block-decimation (TEBD) algorithm \cite{vidal2003,vidal2004,vidal2007,jaschke2018,paeckel2019}. For $0<\alpha<\infty$, we can only benchmark our results against a standard exact diagonalization of the Lindblad generator, possible for relatively small systems. As shown in Fig.~\ref{Fig3}(a-b), the results from our TTN algorithm agree with the corresponding reference solutions. \\

\begin{figure}[t]
\centering
\includegraphics[width=\columnwidth]{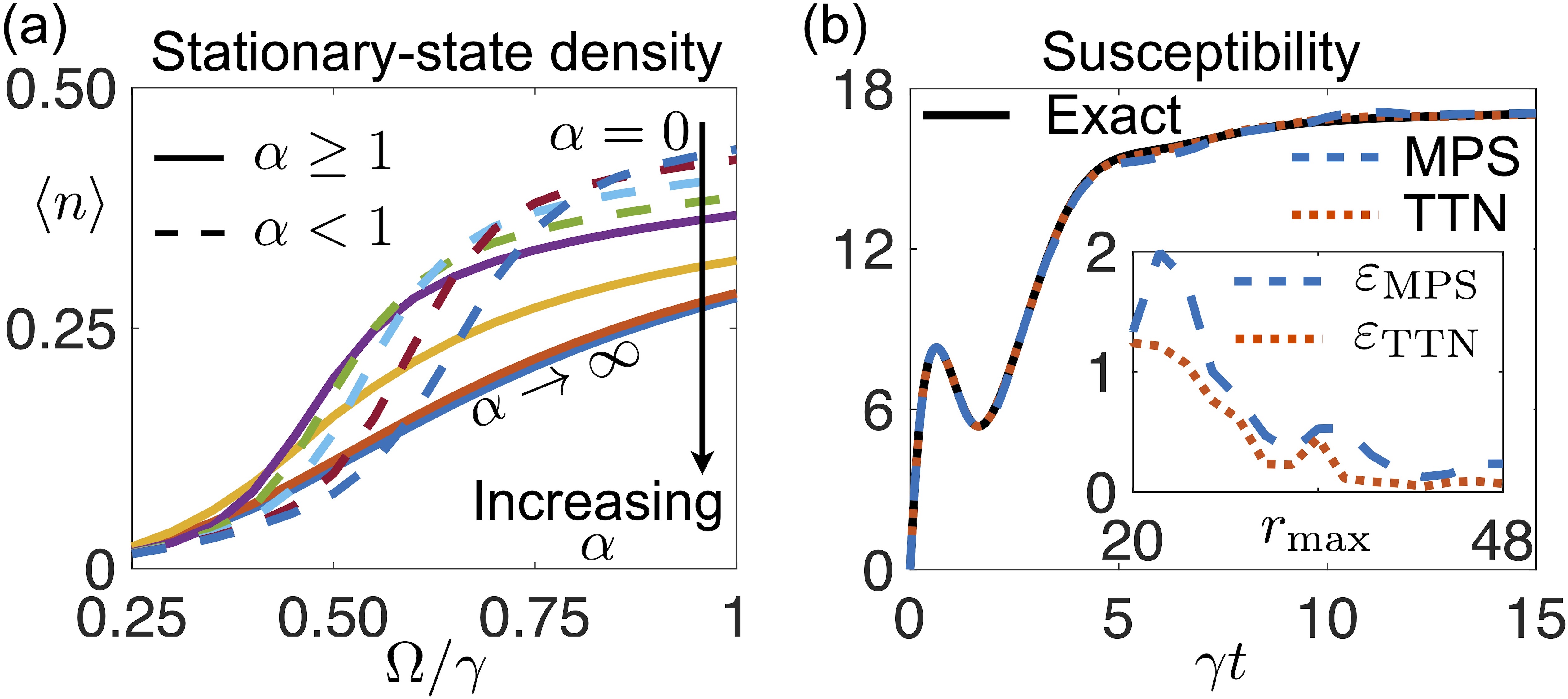}
\caption{{\bf Long-range interactions and ``susceptibility" parameter.} (a) Stationary behavior of the density $\langle n\rangle$ as a function of $\Omega/\gamma$ for different values of $\alpha$ and $D=16$, as estimated at $\gamma t=15$. We consider $\alpha=0,0.25,0.50,0.75,1,2,5$ and $\alpha\to\infty$, growing as denoted in the plot. For all values of $\alpha$, we consider $r_{\rm max}=30$ but for $\alpha=0$, for which $r_{\rm max}=20$ is sufficient. (b) Dynamics of the susceptibility $\chi$ for $\Omega/\gamma=0.65$, for both TTN and MPS simulations with $r_{\rm max}=38$ and $D=32$. The TTN curve essentially coincides with the exact solution while the MPS curve still shows some deviation. The inset displays the maximal errors $\varepsilon_{\rm TTN/MPS}$ in estimating $\chi$ as a function of the bond dimension. The parameters not explicitly specified in the panels are $\Delta/\gamma=-2$, $V/\gamma=5$. }
\label{Fig4}
\end{figure} 

\noindent {\bf Role of the interaction range.---} We now exploit our algorithm to explore the behavior of the system for intermediate values of $\alpha$ and larger system sizes. Such regime is of interest for at least two reasons. First, values such as $\alpha=3$ or $\alpha=6$ are typically encountered in experiments \cite{saffman2010}. Second, for $\alpha=0$ and in the thermodynamic limit $D\to\infty$, the dissipative Ising model features a first-order nonequilibrium phase transition from a phase with a low density of excitations $\langle n\rangle$ to a highly excited one  [cf.~dashed line in Fig.~\ref{Fig3}(b)]. On the other hand, for $\alpha\to\infty$ the transition is not present in the one-dimensional model \cite{jin2018}. Our TTN algorithm can interpolate between these two regimes and allows us to explore the fate of the transition for $\alpha>0$. In Fig.~\ref{Fig4}(a), we see that for $\alpha>1$ the stationary density $\langle n\rangle$ behaves similarly to the case  $\alpha\to\infty$, i.e., there appears to be a smooth behavior of the density $\langle n\rangle$ as a function of $\Omega/\gamma$. On the other hand, for $\alpha\le 1$ for which the sum of the interaction terms in Eq.~\eqref{H_2}, without considering $c_\alpha$, would become super-extensive,  we observe the emergence of a sharp crossover which is reminiscent of what happens in the $\alpha=0$ case. 

To assess the capability of TTNs to capture correlations, we also consider the total density fluctuations in the system $\chi=\sum_{k,h=1}^N\left(\langle n^{(k)}n^{(h)}\rangle -\langle n^{(k)}\rangle \langle n^{(h)}\rangle\right)$. This quantity, which is highly nonlocal as it contains all possible two-body density-density correlations, represents a  susceptibility parameter. Here, we focus on the case $\alpha=0$, for which we can obtain exact results for larger systems \cite{chase2008,baragiola2010,kirton2017,shammah2018}, and compare results for TTNs and MPSs. MPS simulations are also performed using our algorithm  for a TTN of maximal height, which is equivalent to the MPS ansatz \cite{CLS2022}. In Fig.~\ref{Fig4}(b), we display simulations with a same, relatively large, bond dimension for TTNs and MPSs. The plot shows that the TTN results are almost perfectly overlapping with the exact values of the susceptibility, while deviations can still be appreciated in the MPS simulations. In the inset of Fig.~\ref{Fig4}(b), we show the errors $\varepsilon_{\rm TTN/MPS}=\max_{\gamma t\in[0,15]}\left|\chi_{\rm TTN/MPS}-\chi\right|$, where $\chi$ is the exact susceptibility, while $\chi_{\rm TTN/MPS}$ the value estimated with TTNs and MPSs, respectively. Already for the simple all-to-all ($\alpha=0$) interaction considered, which does not develop critical long-range correlations since it features a first-order transition in the thermodynamic limit, we can observe that TTNs perform systematically better than MPSs. More precisely,  we observe, in the inset of Fig.~\ref{Fig4}(b), that TTNs describe more accurately than MPSs the behavior of the susceptibility for a same bond dimension.  \\

\noindent {\bf Discussion.---} We have introduced a method for the numerical simulation of long-range open quantum systems with TTNs and benchmarked it considering the paradigmatic dissipative Ising model. With our method, we could explore the regime of intermediate interaction ranges where we found signatures of the persistence of the phase transition for $\alpha\in[0,1]$, in the thermodynamic limit. We also tested the capability of TTNs to encode correlations. We found that for the considered system, TTNs perform better than simulations with MPSs. Our simulations were performed using standard PCs. 

As a future perspective, it would be interesting to compare the two approaches for open quantum systems featuring second-order nonequilibrium phase transitions and a critical building-up of correlations \cite{carollo2019}. It would also be relevant to explore different tree structures. Here, we mainly considered balanced binary trees and MPSs, but the algorithm is general and applies to any tree  \cite{CLS2022}. This opens up the possibility of a systematic investigation on the role of the tree structure in the encoding of many-body correlations for extended open quantum systems. \\

\acknowledgements
\noindent {\bf Acknowledgements.---} We acknowledge the use of the MATLAB tensor toolbox \cite{TTB_Software} and tensorlab \cite{tensorlab3.0} for the implementation of the algorithm. We acknowledge funding from the German Research Foundation (DFG) through the Research Unit FOR 5413/1, Grant No. 465199066.  FC~is indebted to the Baden-W\"urttemberg Stiftung for financial support by the Elite Programme for Postdocs. The work of GC was supported by the SNSF research project “Fast algorithms from low-rank updates”, grant number 200020-178806.

\bibliography{biblio}

\newpage

\setcounter{equation}{0}
\setcounter{figure}{0}
\setcounter{table}{0}
\makeatletter
\renewcommand{\theequation}{S\arabic{equation}}
\renewcommand{\thefigure}{S\arabic{figure}}

\makeatletter
\renewcommand{\theequation}{S\arabic{equation}}
\renewcommand{\thefigure}{S\arabic{figure}}

\onecolumngrid
\newpage

\setcounter{page}{1}

\begin{center}
{\Large SUPPLEMENTAL MATERIAL}
\end{center}
\begin{center}
\vspace{0.8cm}
{\Large Numerical simulations of long-range open quantum many-body dynamics with tree tensor networks}
\end{center}
\begin{center}
Dominik Sulz$^1$, Christian Lubich$^1$, Gianluca Ceruti$^2$, Igor Lesanovsky$^{3,4}$, and Federico Carollo$^3$
\end{center}

\begin{center}
$^1${\em Mathematisches Institut, Universit\"at T\"ubingen, Auf der Morgenstelle 10, D–72076 T\"ubingen, Germany}\\
$^2$ {\em Institute of Mathematics, EPF Lausanne, 1015 Lausanne, Switzerland}\\
$^3${\em Institut f\"ur Theoretische Physik, Universit\"at T\"ubingen,}\\
{\em Auf der Morgenstelle 14, 72076 T\"ubingen, Germany}\\
$^4${\em School of Physics and Astronomy and Centre for the Mathematics}\\
{\em and Theoretical Physics of Quantum Non-Equilibrium Systems,}\\
{\em The University of Nottingham, Nottingham, NG7 2RD, United Kingdom}
\end{center}

\section*{I. Matrix representation of the Lindblad generator}
We show here how the time evolution of the density matrix $\rho(t)$ can be formulated in terms of a vector differential equation. For the sake of concreteness, we focus on the case of the dissipative Ising model discussed in the main text, for which the time evolution of the density matrix is given by the Lindblad equation
$$
\dot{\rho}(t)=-i[H,\rho(t)]+\sum_{k=1}^D \gamma \left(\sigma_-^{(k)}\rho(t)\sigma_+^{(k)}-\frac{1}{2}\left\{\rho(t), n^{(k)}\right\}\right)\, .
$$
For this model, we have the single-particle basis states $\ket{\bullet},\ket{\circ}$, with which we can define $\sigma_-=\ket{\circ}\!\bra{\bullet}$, $\sigma_+=\sigma_-^\dagger$ and $n=\ket{\bullet}\!\bra{\bullet}$. The system Hamiltonian is 
\begin{equation}
H=\Omega \sum_{k=1}^D\sigma_x^{(k)}+\Delta \sum_{k=1}^D n^{(k)}+\frac{V}{2c_\alpha}\sum_{k\neq h =1}^D \frac{n^{(k)}n^{(h)}}{|k-h|^\alpha}\, ,
\label{}
\end{equation}
with $\sigma_x=\sigma_-+\sigma_+$.

The starting point of the mapping of the above matrix equation into a vector one is to take the density matrix $\rho(t)$, and write it as a vector in an enlarged single-particle Hilbert space. This can be achieved, for instance, through the following mapping 
$$
\rho(t)=\sum_{\vec{\ell},\vec{m}}r_{\vec{\ell}\vec{m}}(t)\ket{\vec{\ell}}\!\bra{\vec{m}}\longrightarrow \ket{\rho(t)}=\sum_{\vec{\ell},\vec{m}}r_{\vec{\ell}\vec{m}}(t)\bigotimes_{k=1}^D \left[\ket{\ell_k}\otimes \ket{m_k}\right]\, .
$$
Here, we have that  $\vec{\ell}=(\ell_1,\ell_2,\dots \ell_D)$ and $\vec{m}=(m_1,m_2,\dots m_D)$ are many-body configuration states, where $\ell_k,m_k=\bullet,\circ$ specify the single-particle state. In this representation, the Lindblad generator  is given  by the following matrix 
\begin{equation}
\begin{split}
\mathbb{L}&=\gamma \sum_{k=1}^D \left(\sigma_{-,{\rm I}}^{(k)}\sigma_{-,{\rm II}}^{(k)}-\frac{1}{2} {n}_{{\rm I}}^{(k)}-\frac{1}{2}{n}_{{\rm II}}^{(k)}\right)-i\sum_{k=1}^{D}\Omega\left( \sigma_{x,{\rm I}}^{(k)} -\sigma_{x,{\rm II}}^{(k)}\right)+\, \\
&- i\sum_{k\neq h}^{D}\frac{V}{2c_\alpha |k-h|^{\alpha}}\left(n_{{\rm I}}^{(k)}n_{,{\rm I}}^{(h)}-n_{{\rm II}}^{(k)}n_{,{\rm II}}^{(h)}\right)
\label{mat-L-rep}
\end{split}
\end{equation}
where $\sigma_{-,{\rm I}}=\sigma_-\otimes {\bf 1}_2$, $\sigma_{-,{\rm II}}={\bf 1}_2\otimes\sigma_-$ and similarly,  $\sigma_{x,{\rm I}}=\sigma_x\otimes {\bf 1}_2$, $\sigma_{x,{\rm II}}={\bf 1}_2\otimes\sigma_x$ as well as $n_{\rm I}=n\otimes {\bf 1}_2$, $n_{{\rm II}}={\bf 1}_2\otimes n$ and ${\bf 1}_2$ is the $2\times2$ identity.  Note that, in principle, one should have transposition of all the terms denoted with ${\rm II}$ in the above Eq.~\eqref{mat-L-rep}, exception made for those in the first term of the first sum (see also, e.g., Refs.~\cite{kshetrimayum2017,carollo2022}). However, in our case all the matrices involved are already self-transposed.  
The time evolution is thus implemented via the vectorized differential equation 
$$
 \ket{\dot{\rho}(t)}=\mathbb{L}\ket{\rho(t)}\, . 
$$

To conclude we recall how expectation values can be computed within this vectorized formalism. Let us consider an elementary operator 
$$
O=\bigotimes_{k=1}^D x_k\, ,
$$
where $x_k$ are $2\times2$ matrices. Then, its expectation value can be  computed as 
$$
\langle  O\rangle_t=\bra{-}\bigotimes_{k=1}^D x^{(k)}_{k,{\rm I}}\ket{\rho(t)}\, ,
$$
where we have defined $x_{k{\rm I}}=x_k\otimes {\bf 1}_2$ as well as the vector representation of the identity
$$
\ket{-}=\bigotimes_{k=1}^D\ket{{\bf 1}_2}\, ,
$$
with $\ket{{\bf 1}_2}=\ket{\bullet}\otimes \ket{\bullet}+\ket{\circ}\otimes \ket{\circ}$. In the main text, we always considered as initial state the state
$$
\ket{\rho(0)}=\bigotimes_{k=1}^D \left(\ket{\bullet}\otimes \ket{\bullet}\right)\, .
$$

\section*{II. Additional details on the tree tensor network algorithm}

\subsection*{A. Recursive definition of a tree tensor network (TTN)}
Suppose to be given a set of basis matrices $U_j$ for $j=1,\dots,D$ and of connecting tensors $C_{\tau_i^j}$ for all subtrees $\tau_i^j$ of $\tau^0$. We recursively define a tree tensor network $X_{\tau^0}$ as follows
\begin{enumerate}
    \item[(i)] For each leaf, we set
    \begin{align*}
            X_j := U_j^T \in \mathbb{C}^{r_j \times n_j}.
    \end{align*}
    \item[(ii)] For each subtree $ \tau_i^j = (\tau_{2i-1}^{j+1},\tau_{2i}^{j+1}) $ of the maximal tree $\tau^0$, we set $n_{\tau_i^j} = n_{\tau_{2i-1}^{j+1}} n_{\tau_{2i}^{j+1}}$ and
    \begin{align*}
        X_{\tau_i^j} &:= C_{\tau_i^j} \times_1 U_{\tau_{2i-1}^{j+1}} \times_2 U_{\tau_{2i}^{j+1}}  \in \mathbb{C}^{r_{\tau_i^j} \times n_{\tau_{2i-1}^{j+1}} \times n_{\tau_{2i}^{j+1}}}, \\
        U_{\tau_i^j} &:= {\rm Mat}_0(X_{\tau_i^j})^T \in \mathbb{C}^{n_{\tau_i^j} \times r_{\tau_i^j}}.
    \end{align*}
\end{enumerate}

\subsection*{B. Construction of the $M_{\tau}$ matrices}
The matrix $M_{\tau}=U_{\tau}^{' \dagger}U_{\tau}$, where $\tau =(\tau_1,\tau_2)$, can be constructed recursively. By definition of a tree tensor network we know that it holds
\begin{align*}
    U_{\tau}^{'} &= {\rm Mat}_0 (C_\tau^{'} \times_1 U_{\tau_1}^{'} \times_2 U_{\tau_2}^{'}  )^T  \\
    U_{\tau}&= {\rm Mat}_0 (C_\tau \times_1 U_{\tau_1} \times_2 U_{\tau_2} )^T  ,
\end{align*}
where $U_{\tau_i}^{'}$ and $U_{\tau_i}$, for $i=1,2$, are either basis matrices or again a matricized tree tensor network from the level below. Using the unfolding formula for tree tensor networks (see equation 2.2 in \cite{CLS2022}) we obtain
\begin{align*}
    M_{\tau}&=U_{\tau}^{' \dagger}U_{\tau} = \left( {\rm Mat}_0 (C_\tau^{'} \times_1 U_{\tau_1}^{'} \times_2 U_{\tau_2}^{'}  )^T \right)^\dagger  {\rm Mat}_0 (C_\tau \times_1 U_{\tau_1} \times_2 U_{\tau_2} )^T \\
    &= \overline{{\rm Mat}_0 (C_\tau^{'})} \left(  \times_1 U_{\tau_1}^{' \dagger}U_{\tau_1} \times_2 U_{\tau_2}^{' \dagger}U_{\tau_2} \right) {\rm Mat}_0 (C_\tau)^T.
\end{align*}
The products $U_{\tau_i}^{' \dagger}U_{\tau_i}$, for $i=1,2$, can now be computed recursively until we reach the basis matrices. 

\subsection*{C. Constructing and applying the tree tensor network operators (TTNOs) ${L}_\tau$} 
As we have shown in the first section of this Supplemental Material, the Lindblad operator can be written, in its matrix representation,  as a linear operator of the form
\begin{equation*}
    \mathbb{A} = \sum_{k=1}^s a_k^1 \otimes \dots \otimes a_k^D,
\end{equation*}
where $a_k^j$ are complex matrices which act on the $j$th particle. Similar ideas for the construction and application of TTNO's can be found in ~\cite{KT2014}. We define the tree tensor network operator $A = {L}_{\tau^0}$, which acts on a tree tensor network $X_{\tau^0}$, to be the tensor network with
\begin{enumerate}
    \item The $j$th leaf equal to the matrix $[\textbf{vec}(a_1^j),\dots,\textbf{vec}(a_s^j)]$, where $\textbf{vec}(B)$ denotes the vectorization of the matrix $B$.
    \item All connecting tensors $C_{\tau} \in \mathbb{C}^{s\times s \times s}$ with entries $C(k_1,k_2,k_3) = 1$ if and only if $k_1=k_2=k_3$. Else the entries are zero.
    \item The connecting tensor $C_{\tau^0} \in \mathbb{C}^{s\times s \times 1}$ at the top level with again entries $C_{\tau^0}(k_1,k_2) = 1$ if and only if $k_1=k_2$, and otherwise zero. 
\end{enumerate}
The resulting tree tensor network should be then orthonormalized and possibly truncated to a reasonable bond dimension. We will call this orthonormal TTN $\Tilde{A}$. Now we define the application of $\Tilde{A}$ to a tree tensor network $X_{\tau^0}$. The leaves and connecting tensors are applied in the following way:
\begin{enumerate}
    \item Let $U_j$ be the $j$th leaf of $X_{\tau^0}$. Then the $j$th leaf of $\Tilde{A}(X_{\tau^0})$ is defined as the matrix $[\Tilde{a}_1^j U_j,\dots,\Tilde{a}_n^j U_j]$, where $\Tilde{a}_i^j$ are the matricizations of the $i$th columns of the $j$th leaf of the TTNO $\Tilde{A}$.
    \item Let $C_{\tau_i^j}$ be the connecting tensor at $j$th level and $i$th position of $X_{\tau^0}$ and respectively $\Tilde{C}_{\tau_i^j}$ the connecting tensor at $j$th level and $i$th position of $\Tilde{A}$. Then the connecting tensor of the application is defined as $\Tilde{C}_{\tau_i^j} \otimes C_{\tau_i^j}$, where $\otimes$ denotes the canonical extension of the Kronecker product to tensors.
\end{enumerate}

\subsubsection*{1. Constructing ${L}_{\tau_i^j}$}
Suppose to be given a TTNO ${L}_{\tau^0}$, constructed as above. Now we are interested in constructing the sub-functions ${L}_{\tau_i^j}$, which are needed for the algorithm (see main text). The definition of these functions is again done recursively from the root to the leaves.

Suppose that for a tree $\tau=(\tau_1,\tau_2)$ the function ${L}_{\tau}$ is already constructed. Let $X_{\tau}$ be a TTN with connecting tensor $C_\tau$ and matrices $U_{\tau_1} = {\rm Mat}_0(X_{\tau_1})^T$ and $U_{\tau_2}={\rm Mat}_0(X_{\tau_2})^T$, i.e. $X_{\tau} = C_\tau \times_1 U_{\tau_1} \times_2 U_{\tau_2}$. We define the space $\mathcal{V}_{\tau} = \mathbb{C}^{r_{\tau} \times n_{\tau_1} \times n_{\tau_2}}$, where $n_{\tau} = \prod_{i=1}^2 n_{\tau_i}$ is defined recursively. We define the matrices
\begin{align*}
    V_{\tau_1}^0 = {\rm Mat}_1 \left( {\rm Ten}_1(Q_{\tau_1}^T) \times_{2} U_{\tau_2} \right)^T, \\
    V_{\tau_2}^0 = {\rm Mat}_2 \left( {\rm Ten}_2(Q_{\tau_2}^T) \times_{1} U_{\tau_1} \right)^T,
\end{align*}
where $Q_{\tau_i}$, for $i=1,2$, is the unitary factor in the QR-decomposition of ${\rm Mat}_0(C_{\tau_i})^T = Q_{\tau_i}R_{\tau_i}$ and $C_{\tau_i}$ is the connecting tensor of $X_{\tau_i}$. Further we define two functions
\begin{align*}
    \pi_{\tau,i} (Y_{\tau_i}) &= {\rm Ten}_i((V_{\tau_i}^0 {\rm Mat}_0(Y_{\tau_i}))^T) \in \mathcal{V}_{\tau}, \ \text{for} \ Y_{\tau_i} \in \mathcal{V}_{\tau_i} \\
     \pi_{\tau,i}^\dagger (Z_{\tau}) &= {\rm Ten}_0 (({\rm Mat}_i(Z_{\tau})V_{\tau_i}^0 )^T) \in \mathcal{V}_{\tau_i}, \ \text{for} \ Z_{\tau} \in \mathcal{V}_{\tau}.
\end{align*}
The function ${L}_{\tau_i}$ now is defined recursively by
\begin{equation}
    {L}_{\tau_i} = \pi_{\tau,i}^\dagger \circ {L}_{\tau} \circ \pi_{\tau,i}, \ \ \text{for} \ i=1,2.
    \label{function_F}
\end{equation}

\end{document}